\RequirePackage{fix-cm}
\documentclass[twocolumn]{article}

\usepackage[a4paper, margin=0.5in]{geometry}
\usepackage{graphicx}
\usepackage{calc}
\usepackage[table]{xcolor}
\usepackage{authblk}

\def\keywords#1{\\\\\textbf{Keywords}. #1.}
\begin{document}

\author[1]{Mohammad Muzammil Khan\footnote{Contact: mmkhan.sch@jamiahamdard.ac.in}}
\author[1]{Shahab Saquib Sohail}
\author[2]{Darakhshan Ishrat}

\affil[1]{Dept. of CSE, SEST, Jamia Hamdard (India)}
\affil[2]{Dept. of IS, SHSS, Jamia Hamdard (India)}

\title{How They Tweet? An Insightful Analysis of Twitter Handles of Saudi Arabia}

\date{}

\maketitle
\begin{abstract}
    The emergence of social network site has attracted many users across the world to share their feeling, news, achievements and personal thoughts over several platforms. The recent crisis due to worldwide lockdown amid COVID 19 has shown how these online social platforms have grown stronger and turned up as the major source of connection among people when there is social distancing everywhere. Therefore, we have surveyed Twitter users and their mannerism with respect to languages, frequency of tweets, the region of belonging, etc. The above observations have been considered especially with respect to Saudi Arabia. An insightful analysis of the tweets and twitter handles of the kingdom has been presented. The results show some interesting facts that are envisaged to lay a platform for further research in the field of social, political and data sciences related to the Middle East.
\end{abstract}

\section{Introduction}
Twitter is a social networking and micro-blogging platform. The service gained rapid success and popularity as it provided a unique way to interact with the Internet termed as “tweets”. Tweets can be a variety of services provided on the platform which include text, photos, videos, polls, etc. The service was launched in the year 2006 and after just 6 years, in 2012, the micro-blogging site had gained 140 million active users with over 300 million tweets a day [1]. The platform initially restricted a tweet with a limit of 140 characters. However, in 2017 Twitter doubled the limit for non-Asian languages to 280 characters [2]. Twitter provides many ways to categorize and tag tweets. For combining similar tweets based on content-type or combining similar tweets based on a ticker symbol for stocks and companies, “hashtags”, can be used and usually known as trend/trending over Twitter.

Twitter generates an enormous amount of data that can be accessed using the official Application Program Interface (API). The API requires an authentication key to access the data. However, there are limitations for each type of API. These limitations manage the number of times a request can be made, the amount of data a single user can access, and how much of the Twittersphere is accessible to a user. These limitations are here to protect data usage and prevent personalized and targeted approach for an individual or a group of individuals.

We aim to collect the data as much as possible respecting the limits and restrictions in place and perform an analysis of the tweets to find some useful analytical data which can be used to analyze the behaviour of users tweeting in a certain region. We shall collect and analyze the tweets from the perspective of the Kingdom of Saudi Arabia (KSA).
\section{Background}
\subsection{Related work}
In general, researchers have been studying and exploring the social network for quite a while now. Many of the researchers have applied their studies and found influence patterns [3]. Researchers, regarding the influencers, have found that user accounts of news media are better spreaders of contents and that the celebrities use Twitter to make conversations [4]. Researchers in [5] focused a study for the United States of America of demographic data using the Twitter data from [3] and have been able to identify a location, gender, and race/ethnicity only from the publicly accessible data. Researchers have been able to use the data generated from online social networking sites for predicting many things that seem random and/or can’t be humanly predicted. These predictions include the stock market [6, 7], book sales [8], movie-ticket sales [9, 10], product sales [11], infections, diseases, and consumer-spending [12]. Authors have also incorporated Twitter data for healthcare purpose and disease prediction [13, 14].

There have been few works with respect to Saudi Arabia and Arabic (language) tweets. In 2011, Al-Khalifa [15] performed a social network analysis of the Twittersphere to understand the social structure of citizens and their interactions as discussions of issues regarding politics. Since then Arabic dialects are receiving a lot of attention from researchers. With the use of natural language processing, many researchers are now able to investigate even more deeply in an unprecedented manner. In [16], Al-Twairesh et al. have conducted a study of the Modern Standard Arabic (MSA) usage by the Arabic-speaking users of Twitter. Just like the predictions made in [6 – 12], researchers have also been able to predict the Saudi Stock Market Index [17]. Authors have also been able to detect sarcasm through analysis of the Arabic tweets [18]. In 2018, researchers targeted the Arabic-tweeting community for profiling to help identify and extract attributes of a Twitter user using machine learning-based classification of topics from Tweets which achieves a $90\%$ accuracy [19]. Sentiment analysis is also being used in research to provide a better understanding of using text mining [20]. An emotion and mood visualizer called “Saudi Mood” was proposed in [21] to continuously monitor the Arabic Twitter to detect dominant emotion in the Kingdom of Saudi Arabia using real-time sentiment analysis. Another sentiment analysis [22] was used to get an insight into the different dialects used in the Arabic tweets using sentiment analysis of the Arabic tweets.

\subsection{Key issues addressed in this research}
As of January 2020 [23], there are 14.35 million Twitter users from Saudi Arabia, with a rank of 4, followed by 59.35, 45.75, and 16.7 million users from the United States, Japan, and the United Kingdom respectively. As the number of users is higher, we can infer that the number of tweets from the Kingdom is higher as well. We have investigated whether Saudi Arabia falls in the top 10 most tweeting countries or not?

In addition, Arabic is the lingua franca of the Arab world. Being the 6th most-spoken language worldwide [24], we have critically reconnoitred the ranking of Arabic among the top 10 languages used on Twitter. Regardless of the dialects spoken, Egypt is the most populous Arabic-speaking country [25]. However, a lot of factors come into play like social awareness of the platform, social networking norm in the country, etc. when we ponder over who tweets more. According to [26], Facebook is the most popular social media platform used by the Egyptian public and Twitter is the 3rd. According to a report [27], Facebook is also the most favoured in Saudi Arabia, however, Twitter comes 2nd. Naturally, we have inspected that which country will have more Arabic-tweets, Saudi Arabia, or any other country like Egypt?

A study conducted in 2016 [28] shows that the best time to get most interactions for a tweet is around noon, from 12 PM to 1 PM. This study was conducted on a high volume of tweets. However, they did not include or mention Saudi Arabia in the research explicitly. We have surveyed the Twitter data to find out the period in which most frequent tweets are tweeted concerning Saudi Arabia.

\section{Data crawling and information fetching}

To maximize efficiency and productivity, we have divided the work into different yet integrated modules. The whole set consisted of 4 modules which performed tasks by taking the input of the previous module's data and generate output to be used as input for the next module. The modules are – Crawler, Processor, Analyzer, and Pruner.

A customized crawler was created which accessed the tweets using the official API in real-time and stored the data synchronously in a well-defined format. We ran the crawler for around 23 continuous days with API rate-limiting of 450 requests per 15 minutes and got over 86.79 million tweets. The Crawler was programmed to save only the tweets in which – 1. the user has entered their location, and 2. language was detected by Twitter. After trimming the results, we had a dataset consisting of 58.15 million tweets.  We also created a custom processor to process in whichever way we wanted. In the Processor, we programmed two sub-modules, i.e. a sub-module to check and verify the integrity of dataset and a sub-module to detect and extract countries and cities from a plain text which in this case is the location field self-reported by the user. We ran the Processor twice. In the first time, we performed a check on duplication and the false negatives of location detection. After the first round of processing, we found that there were 8.2 million duplicate tweets and 2.36 million distinct entries for unknown/undetected locations from 21.69 million tweets. Then, we manually evaluated the false negatives and adjusted the algorithm with a modification of 0.37 million entries. While evaluating, we found that people were using a wide range of spellings and native languages for their locations. This process also confirms that the dataset format is error-free as we were able to process the whole dataset of crawled tweets. 

After deduplication and adjustments, we ran the Processor again, we had a data-set of 49.99 million tweets and were able to detect countries of 68.49\% of the data-set. Then the analyzer came into the picture. The analyzer analyzed each tweet depending upon the inputs provided to categorize the data. We used a combination of regular expressions, mapping, and hardcoded commands to sort out the data and to output the analysis in comma-separated-value format. This format can be imported into any analysis tool for further study. We programmed the Analyzer to show some of the preliminary analysis after the completion of execution. The preliminary analysis showed that there were – 12.85 million users from which 7.24 million posted tweets and 8.67 retweeted with 3.07 million users doing both, the dataset had 33.84 million words, 64 languages, and 237 countries. Finally, the Pruner analyzed the data which was more than 16 GB and was too big to load every time to look upon. So, the Pruner produced the output, i.e. the sorted and trimmed data, which was relatively small and much portable.

\begin{table*}
\caption{Languages used for tweeting in KSA}
\label{tab:languages}
\begin{tabular}{llllllll}
\hline\noalign{\smallskip}
Language & Tweets & Language & Tweets & Language & Tweets & Language & Tweets \\
\noalign{\smallskip}\hline\noalign{\smallskip}
Arabic & 61390 & Korean & 75 & German & 19 & Norwegian, Malayalam & 4 \\
English & 27161 & Hindi & 66 & Romanian & 14 & Thai, Swedish, Danish & 2 \\
Spanish & 464 & Persian & 59 & Vietnamese & 13 & Bangla, Kannada, \\
Catalan & 407 & Italian & 55 & Welsh & 11 & Polish, Telugu, \\
Urdu & 403 & Japan & 53 & Czech & 10 & Sindhi, Hebrew, & 1 \\
Tagalog & 396 & Haitian & 38 & Tamil & 9 & Icelandic, Lithuanian \\
French & 267 & Turkish & 31 & Finnish, Chinese & 8 \\
Portuguese & 226 & Estonian & 25 & Dutch & 7 \\
Indonesian & 161 & Pashto & 20 & Hungarian & 5 \\
\noalign{\smallskip}\hline
\end{tabular}
\end{table*}

\begin{table}
\caption{Arabic-tweeting countries}
\label{tab:countries}
\begin{tabular}{llllllll}
\hline\noalign{\smallskip}
Rank & ISO 3166-2 & Country & Tweets \\
\noalign{\smallskip}\hline\noalign{\smallskip}
- & - & Unknown countries & 98707 \\
1 & SA & Saudi Arabia & 61390 \\
2 & EG & Egypt & 12415 \\
3 & KW & Kuwait & 9306 \\
4 & AE & United Arab Emirates & 4995 \\
5 & US & United States & 3748 \\
6 & IQ & Iraq & 2374 \\
7 & BH & Bahrain & 1711 \\
8 & OM & Oman & 1609 \\
9 & QA & Qatar & 1592 \\
10 & GB & United Kingdom & 1274 \\
- & - & Others & 10122 \\

\noalign{\smallskip}\hline
\end{tabular}
\end{table}

\section{Results and Discussion}
\subsection{Frequency of the Tweets}

\begin{figure}
	\includegraphics[width=0.49\textwidth]{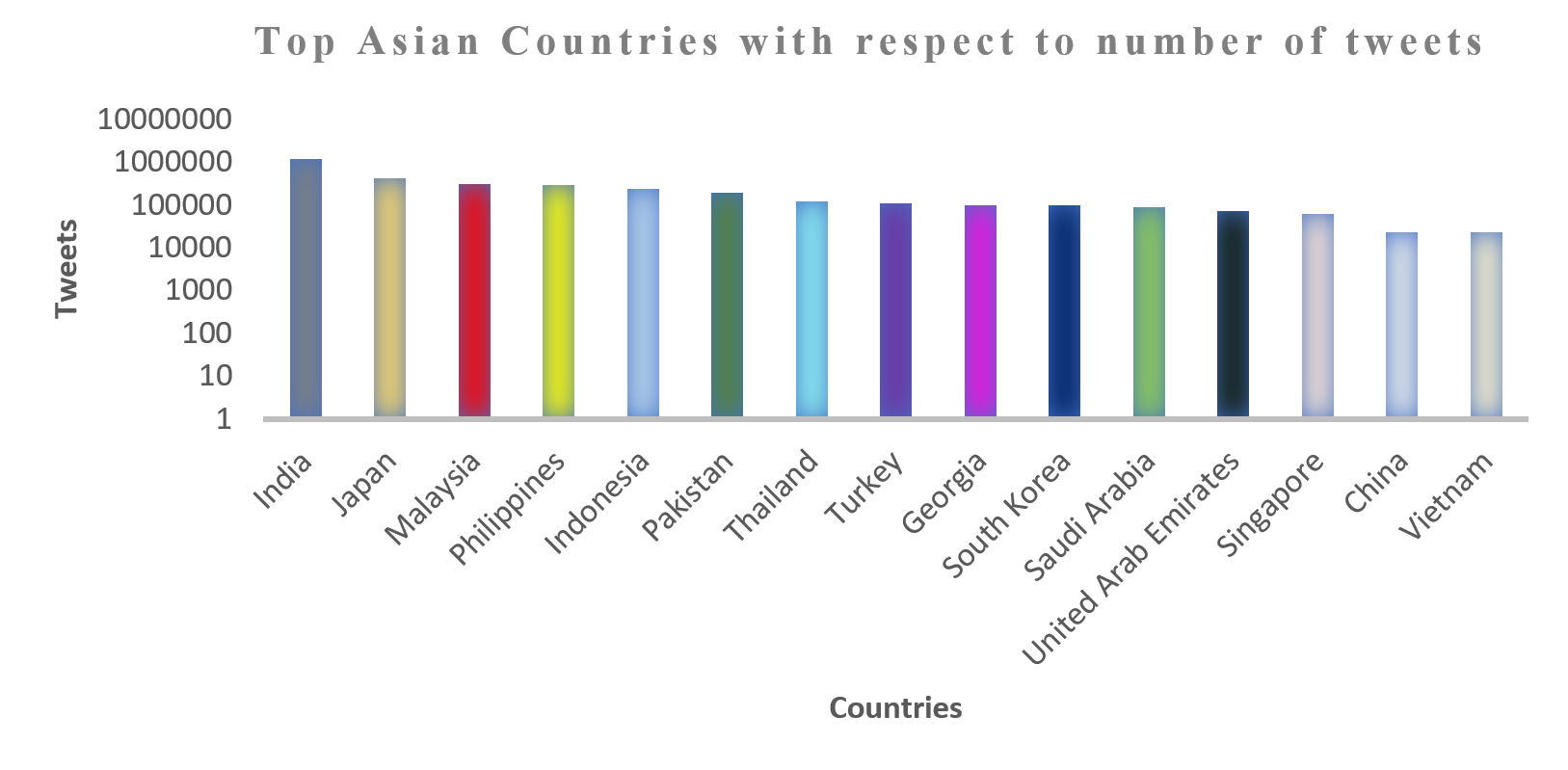}
	\caption{Asian countries and number of tweets recorded during the period of the experiment}
	\label{fig:countries}
\end{figure}

Our analysis of the crawled tweets has shown that among the 237 countries, the United States of America was on the top in the frequency of tweet creation, retweeting, and both combined, followed by Brazil in each category. The Kingdom of Saudi Arabia ranked 27th in tweet creation, 37th in retweeting, and 35th in the overall frequency of tweets. On a continental level, The Kingdom of Saudi Arabia is ranked at 11th position with India having ranked first as shown in figure \ref{fig:countries}.

\begin{figure}
	\includegraphics[width=0.49\textwidth]{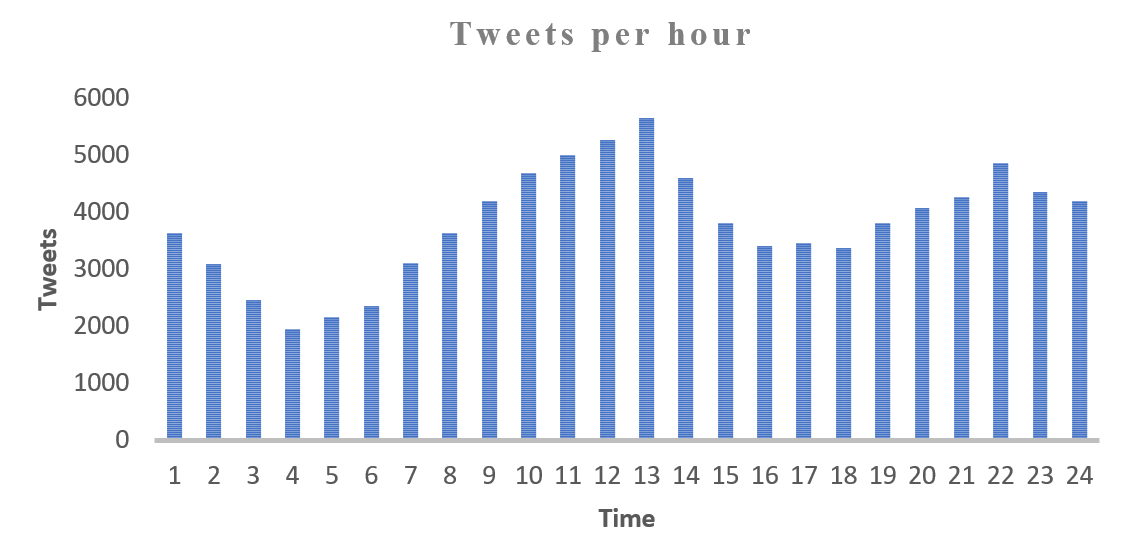}
	\caption{Time duration and total tweeted Tweets during the period}
	\label{fig:time}
\end{figure}

\subsection{Most frequent language in tweets}
Twitter provides with language detection of tweets within the API. At the time of crawling, we selected and saved only the tweets in which the API was able to detect the language. Our dataset had a total of 64 languages, among which English was the most popular language with 29.35 million tweets, followed by Spanish with 8.62 million and Portuguese at 6.31 million, and the least was, with mere 5 tweets, Laotian – a native language of the people of Laos.

Here, we infer that Arabic which is the official language of the Kingdom, ranked at 8th position with over 0.21 million tweets or amounting to 0.42\% of the whole dataset. Moreover, as shown in table \ref{tab:languages}, Arabic is found to be the most popular language in the KSA with 61,390 tweets, amounting to 67.15\% of tweets originating from the country. Followed by English and Spanish with 27,161 and 464 tweets respectively and Lithuanian, Icelandic, Hebrew, Sindhi, Telugu, Polish, Kannada, and Bangla languages having only 1 tweet each.

Despite being the most popular language among the user base of Twitter residing in the Kingdom, Arabic tweets from the Kingdom only amount to 29.34\% of the whole dataset and 47.17\% of tweets come from unidentified locations from which a percentage could very well be residing in the KSA itself. The KSA is followed by Egypt, Kuwait, the United Arab Emirates, the United States, and Iraq with 5.93\%, 4.45\%, 2.39\%, 1.79\%, and 1.13\% of the Arabic tweets respectively (table \ref{tab:countries}).

\subsection{Usage during the day}
The Twitter API returned each tweet with the timestamp of creation in the Coordinated Universal Time (UTC) format. We stored this as well and then converted the format into Arabian Standard Time (AST). We saw that the Saudis are the least active at dawn, specifically 0400 hours AST or 0100 UTC, slowly picking pace and achieving peak usage in the afternoon, specifically 1300 hours AST or 1000 hours UTC. Afterwards, usage declines until 1600 hours AST (1300 UTC) to 1800 hours AST (1500 UTC) (figure \ref{fig:time}).

This finding corroborates with the study [28] and suggests that the best time to get most interactions from a tweet is at noon, specifically around 1300 AST, as most of the users are active around that time.

\section{Conclusion}

In this paper, we crawled around 86 million tweets during 23 days and then processed and analyzed the data concerning the trends and behaviour in Saudi Arabia and the Arabic language over Twitter. Primarily we are focused on exploring whether Saudi Arabia is in the top 10 tweeting countries as the number of Twitter users in the KSA is relatively higher, hence more users should result in more tweets. We have found that Saudi Arabia is placed in the 11th position as far as the number of tweets is concerned. Since Arabic is among the leading languages spoken worldwide, therefore, we have also identified how users at Twitter across the Globe behave when language is concerned and it is found that Arabic remains one of the top languages with respect to its frequency over Twitter.  In addition, it has been observed that the Kingdom of Saudi Arabia has the most Arabic-tweets when comparing to Arabic tweets from any other country worldwide. The earlier study suggests that the most active time for users is at noon. We have investigated and found the most frequent tweeting time comes to be 12 pm – 1 pm across the respective time zones of the countries concerned. Further, in the future, we would like to analyze this area by performing deep analysis on the tweets themselves and their impact on the Arab culture. We can perform sentiment analysis on tweets to further classify and extract features. This study is believed to establish a base of future Twitter research not only for Arab-centric but also to other regions and other aspects.


\begin{thebibliography}{}
\bibitem{} Twitter turns six. https://blog.twitter.com/official/en\_us/\\a/2012/twitter-turns-six.html. Accessed on 5 May, 2020.
\bibitem{} Tweeting Made Easier. https://blog.twitter.com/official\\/en\_us/topics/product/2017/tweetingmadeeasier.html. Accessed on 5 May, 2020.
\bibitem{} Cha, M., Haddadi, H., Benevenuto, F., \& Gummadi, K. P. (2010, May). Measuring user influence in twitter: The million follower fallacy. In fourth international AAAI conference on weblogs and social media.
\bibitem{} Leavitt, A., Burchard, E., Fisher, D., \& Gilbert, S. (2009). The influentials: New approaches for analyzing influence on twitter. Web Ecology Project, 4(2), 1-18.
\bibitem{} Mislove, A., Lehmann, S., Ahn, Y. Y., Onnela, J. P., \& Rosenquist, J. N. (2011, July). Understanding the demographics of Twitter users. In Fifth international AAAI conference on weblogs and social media.
\bibitem{} Bollen, J., Mao, H., \& Zeng, X. (2011). Twitter mood predicts the stock market. Journal of computational science, 2(1), 1-8.
\bibitem{} Schumaker, R. P., \& Chen, H. (2009). Textual analysis of stock market prediction using breaking financial news: The AZFin text system. ACM Transactions on Information Systems (TOIS), 27(2), 1-19.
\bibitem{} Gruhl, D., Guha, R., Kumar, R., Novak, J., \& Tomkins, A. (2005, August). The predictive power of online chatter. In Proceedings of the eleventh ACM SIGKDD international conference on Knowledge discovery in data mining (pp. 78-87).
\bibitem{} Mishne, G., \& De Rijke, M. (2006, March). Capturing Global Mood Levels using Blog Posts. In AAAI spring symposium: computational approaches to analyzing weblogs (Vol. 6, pp. 145-152).
\bibitem{} Asur, S., \& Huberman, B. A. (2010, August). Predicting the future with social media. In 2010 IEEE/WIC/ACM international conference on web intelligence and intelligent agent technology (Vol. 1, pp. 492-499). IEEE.
\bibitem{} Liu, Y., Huang, X., An, A., \& Yu, X. (2007, July). ARSA: a sentiment-aware model for predicting sales performance using blogs. In Proceedings of the 30th annual international ACM SIGIR conference on Research and development in information retrieval (pp. 607-614).
\bibitem{} Choi, H., \& Varian, H. (2012). Predicting the present with Google Trends. Economic record, 88, 2-9.
\bibitem{} Sadah, S. A., Shahbazi, M., Wiley, M. T., \& Hristidis, V. (2016). Demographic-based content analysis of web-based health-related social media. Journal of medical Internet research, 18(6), e148.
\bibitem{} Udayakumar, S., Senadeera, D. C., Yamunarani, S., \& Cheon, N. J. (2018). Demographics analysis of twitter users who tweeted on psychological articles and tweets analysis. Procedia computer science, 144, 96-104.
\bibitem{} Al-Khalifa, H. S. (2011, December). Exploring political activities in the Saudi Twitterverse. In Proceedings of the 13th International Conference on Information Integration and Web-based Applications and Services (pp. 363-366).
\bibitem{} Al-Twairesh, N., Al-Khalifa, H., \& Al-Salman, A. (2015, April). Towards analyzing Saudi tweets. In 2015 First International Conference on Arabic Computational Linguistics (ACLing) (pp. 114-117). IEEE.
\bibitem{} Hamed, A. R., Qiu, R., \& Li, D. (2015, December). Analysis of the relationship between Saudi twitter posts and the Saudi stock market. In 2015 IEEE Seventh International Conference on Intelligent Computing and Information Systems (ICICIS) (pp. 660-665). IEEE.
\bibitem{} Al-Ghadhban, D., Alnkhilan, E., Tatwany, L., \& Alrazgan, M. (2017, May). Arabic sarcasm detection in Twitter. In 2017 International Conference on Engineering \& MIS (ICEMIS) (pp. 1-7). IEEE.
\bibitem{} Alhozaimi, A., \& Almishari, M. (2018, April). Arabic Twitter Profiling For Arabic-Speaking Users. In 2018 21st Saudi Computer Society National Computer Conference (NCC) (pp. 1-6). IEEE.
\bibitem{} Abo, M. E. M., Raj, R. G., Qazi, A., \& Zakari, A. (2019). Sentiment Analysis for Arabic in Social Media Network: A Systematic Mapping Study. arXiv preprint arXiv:1911.05483.
\bibitem{} Almanie, T., Aldayel, A., Alkanhal, G., Alesmail, L., Almutlaq, M., \& Althunayan, R. (2018, April). Saudi Mood: A Real-Time Informative Tool for Visualizing Emotions in Saudi Arabia Using Twitter. In 2018 21st Saudi Computer Society National Computer Conference (NCC) (pp. 1-6). IEEE.
\bibitem{} Alotaibi, S., Mehmood, R., \& Katib, I. (2019, June). Sentiment Analysis of Arabic Tweets in Smart Cities: A Review of Saudi Dialect. In 2019 Fourth International Conference on Fog and Mobile Edge Computing (FMEC) (pp. 330-335). IEEE.
\bibitem{} Leading countries based on number of Twitter users as of April 2020. https://www.statista.com/statistics/242606/number-of-active-twitter-users-in-selected-countries/. Accessed on 5 May 2020.
\bibitem{} The most spoken languages worldwide in 2019. https://www.statista.com/statistics/266808/the-most-spoken-languages-worldwide/. Accessed on 5 May 2020.
\bibitem{} List of countries where Arabic is an official language. https://en.wikipedia.org/wiki/List\_of\_countries\_where\\\_Arabic\_is\_an\_official\_language. Accessed on 5 May 2020.
\bibitem{} Social Media Stats Egypt. https://gs.statcounter.com/social-media-stats/all/egypt/2019. Accessed on 5 May 2020.
\bibitem{} Social Media Stats Saudi Arabia. https://gs.statcounter.com/social-media-stats/all/saudi-arabia/2019. Accessed on 5 May 2020.
\bibitem{} The Biggest Social Media Science Study: What 4.8 Million Tweets Say About the Best Time to Tweet. https://buffer.com/resources/best-time-to-tweet-research. Accessed on 5 May 2020.

\end{thebibliography}
\end{document}